%% file: main.tex
\documentclass[conference]{IEEEtran}
\IEEEoverridecommandlockouts

\usepackage{cite}
\usepackage{amsmath,amssymb,amsfonts}
\usepackage{algorithmic}
\usepackage{graphicx}
\usepackage{textcomp}
\usepackage{xcolor}
\usepackage{bm}
\usepackage{algorithm}
\usepackage{array}
\usepackage[caption=false,font=normalsize,labelfont=sf,textfont=sf]{subfig}
\usepackage{stfloats}
\usepackage{url}
\usepackage{verbatim}
\usepackage{cite}
\usepackage{multirow,tabularx,enumitem,makecell,booktabs}
\usepackage{lipsum}
\usepackage{arydshln}
\usepackage{breqn}
\usepackage{pifont}%
\newcommand{\cmark}{\ding{51}}%
\newcommand{\xmark}{\ding{55}}%

\def\BibTeX{{\rm B\kern-.05em{\sc i\kern-.025em b}\kern-.08em
    T\kern-.1667em\lower.7ex\hbox{E}\kern-.125emX}}
\begin{document}

\title{Trainable Adaptive Score Normalization for Automatic Speaker Verification}

\author{\IEEEauthorblockN{Jeong-Hwan Choi \dag \qquad Ju-Seok Seong  \qquad Ye-Rin Jeoung \qquad Joon-Hyuk Chang \ddag
\thanks{\dag \, Jeong-Hwan Choi conducted this work during his graduate studies. He is currently working at Samsung Research.} \thanks{\ddag \, Joon-Hyuk Chang is the corresponding author.}
}
\IEEEauthorblockA{\textit{Department of Electronics Engineering, Hanyang University, Seoul, Republic of Korea} \\
\{brent1104, as2835510, jyr0328, jchang\}@hanyang.ac.kr}
}

\maketitle

\begin{abstract}
Adaptive S-norm (AS-norm) calibrates automatic speaker verification (ASV) scores by normalizing them utilize the scores of impostors which are similar to the input speaker.
However, AS-norm does not involve any learning process, limiting its ability to provide appropriate regularization strength for various evaluation utterances.  
To address this limitation, we propose a trainable AS-norm (TAS-norm) that leverages learnable impostor embeddings (LIEs), which are used to compose the cohort.
These LIEs are initialized to represent each speaker in a training dataset consisting of impostor speakers.
Subsequently, LIEs are fine-tuned by simulating an ASV evaluation. 
We utilize a margin penalty during top-scoring IEs selection in fine-tuning to prevent non-impostor speakers from being selected.  
In our experiments with ECAPA-TDNN, the proposed TAS-norm observed 4.11\% and 10.62\% relative improvement in equal error rate and minimum detection cost function, respectively, on VoxCeleb1-O trial compared with standard AS-norm without using proposed LIEs. 
We further validated the effectiveness of the TAS-norm on additional ASV datasets comprising Persian and Chinese, demonstrating its robustness across different languages.
\end{abstract}


\vspace{-0.2cm}
\section{Introduction}
In automatic speaker verification (ASV) systems, score normalization plays a crucial role in generating well-calibrated verification scores. This back-end process effectively compensates for disparities in target and non-target score distributions arising from varied speaker embedding models. In addition, it addresses score fluctuations caused by various test utterance conditions, including the recording channel, acoustic environments, and linguistic content. The significance of score normalization has been recognized since the inception of ASV research \cite{Reynolds1, socre_norm1, socre_norm2}, as it mitigates the challenge of setting thresholds amidst score imbalances caused by various factors.
Score normalization typically involves shift and scale calculations, often exploiting mean and standard deviation (STD) of scores between a set of impostor embeddings (IEs), known as a cohort, and evaluation embeddings extracted from the enrollment and test utterances. 
Over time, various methodologies have been proposed as score normalization technique such as 
Z-norm \cite{z-norm}, T-norm \cite{t-norm}, and their variants \cite{zt-norm, kl-t-norm, s-norm}.
A notable advancement came with the introduction of an adaptive algorithms \cite{at-norm, top-norm, as-norm1, as-norm2, am-norm} that select the top-scoring IEs to compose the cohort, rather than use the entire IEs.
In particular, adaptive S-norm (AS-norm) \cite{as-norm1, as-norm2} has been widely used in various study \cite{am-norm, ad-norm, ecapa-tdnn, bc_cmt, NeXt-TDNN, bc-cmt-kd} including ASV challenges \cite{idlab-SdSV2020, dku-SdSV2021, spkin_ffsvc2022, voxsrc22_hyu} until recently.
However, lack of a learning process in adaptive score normalization raises concerns of the sub-optimal nature of shift and scale computation for score calibration, potentially limiting its calibration ability for diverse evaluation utterances.
In our previous study \cite{sp_cup24_hyu} on IEEE SP cup 2024 \cite{sp_cup24}, we presented the feasibility of trainable AS-norm (TAS-norm), which adjust the shift and scale of the AS-norm by utilizing learnable parameter and in-domain training data, which has similar domain knowledge to the evaluation data. 
In \cite{sp_cup24_hyu}, we introduced learnable weights and biases (LWBs).
However, we have found that our previous TAS-norm only works on the specific dataset.
To address this limitation, we propose a novel TAS-norm that utilizes learnable IEs (LIEs) instead of utilizing LWBs. 
The initial values of these LIEs are set to represent each speaker of the training dataset, equal to the standard AS-norm.
Different from the standard AS-norm, which does not learn IEs, we fine-tune LIEs.
The fine-tuning process for LIEs encompasses the simulation of ASV evaluation using the AS-norm with paired training utterances. 
During the fine-tuning, the LIEs selected for cohort can be non-imposters. To prevent it, we penalize the scores between training utterance's embeddings and LIEs.
We use log-likelihood-ratio cost (CLLR) function \cite{cllr, cllr_loss, wespeaker} to optimize LIEs.
Upon completion of training, the fine-tuned LIEs are utilized for constructing cohort during evaluation.
With this refined cohort, we can get a well-calibrated score that better fits the in-domain data.
To further enhance our proposed method, we introduce two key improvements: a sub-center methodology to boost ASV performance and an auxiliary impostor classification (AIC) loss for stable learning.
Our studies pioneer deep learning approaches in score normalization, enhancing AS-norm across diverse datasets.
To distinguish from our previous study \cite{sp_cup24_hyu}, the previous and proposed versions of the TAS-norm are denoted as \textbf{LWB-TAS-norm} and \textbf{LIE-TAS-norm}, respectively.

\vspace{-0.2cm}
\section{Adaptive score normalization}
Let ($\mathbf{e, t}$) represent a trial comprise of an enrollment embedding $\mathbf{e}$ and test embedding $\mathbf{t}$. These $V$ size vectors, ($\mathbf{e, t}$) are typically extracted from the deep neural network-based speaker embedding model \cite{ecapa-tdnn, bc_cmt, NeXt-TDNN}. The $s(\mathbf{e, t})$ is the score between two embeddings. 
To construct the cohort, let $\mathbf{I} \in \mathcal{R}^{C \times V}$ be the set of IEs representing each impostor's speaker identity, where $C$ is the number of impostors.
Traditional score normalization methods \cite{z-norm, t-norm, zt-norm, kl-t-norm} use all IEs as a cohort.
For example, Z-norm \cite{z-norm} and T-norm \cite{t-norm} employ impostor score distribution obtained by scoring enrollment and test embeddings with all IEs, respectively.
An average of normalized scores obtained from the Z-norm and T-norm is denoted as S-norm \cite{s-norm}, and it is formulated as follows:
\begingroup\makeatletter\def\f@size{9.5}\check@mathfonts
\begin{equation}
\label{eqn:s_norm}
    s_{s}(\mathbf{e, t}) =\frac{1}{2} 
    \left(\frac{s(\mathbf{e, t}) - \mu(\mathbf{s(\mathbf{e, I})})}{\sigma(\mathbf{s(\mathbf{e, I})})} + 
    \frac{s(\mathbf{e, t}) - \mu(\mathbf{s(\mathbf{t, I})})}{\sigma(\mathbf{s(\mathbf{t, I})})} \right),
\end{equation}
\endgroup
where $\mu(\cdot)$ and $\sigma(\cdot)$ denote computing operation for mean and STD, respectively.
The $\mathbf{s}(\mathbf{e, I})$ and $\mathbf{s}(\mathbf{t, I})$ represent the score sets obtained by measuring $\mathbf{e}$ and $\mathbf{t}$ against $\mathbf{I}$, respectively. These score sets are $C$ size vector.
After that, S-norm was extended to incorporate an adaptive cohort selection step, drawing inspiration from adaptive T-norm \cite{at-norm} and Top-norm \cite{top-norm}. This extended version is known as AS-norm \cite{as-norm1, as-norm2}.
In AS-norm, given a cohort size $K$, embedding dependent cohorts $\mathbf{I}_{e} \in \mathcal{R}^{K \times V}$ and $\mathbf{I}_{t} \in \mathcal{R}^{K \times V}$ are constructed by selecting the IEs that yield the highest scores based on $s(\mathbf{e, I})$ and $s(\mathbf{t, I})$, respectively, where $K \leq C$.
Two variants of AS-norm were introduced (AS-norm1 and AS-norm2): The normalized scores of two variants are formulated as follows:
\begingroup\makeatletter\def\f@size{9.5}\check@mathfonts
\begin{equation}
\label{eqn:AS_norm1}
    s_{as1}(\mathbf{e, t}) =
    \frac{s(\mathbf{e, t}) - \mu(\mathbf{s(\mathbf{e, I_{e}})})}{2\,\sigma(\mathbf{s(\mathbf{e, I_{e}})})} + 
    \frac{s(\mathbf{e, t}) - \mu(\mathbf{s(\mathbf{t, I_{t}})})}{2\,\sigma(\mathbf{s(\mathbf{t, I_{t}})})},
\end{equation}
\endgroup
\begingroup\makeatletter\def\f@size{9.5}\check@mathfonts
\begin{equation}
\label{eqn:AS_norm2}
    s_{as2}(\mathbf{e, t}) =
    \frac{s(\mathbf{e, t}) - \mu(\mathbf{s(\mathbf{e, I_{t}})})}{2\,\sigma(\mathbf{s(\mathbf{e, I_{t}})})} + 
    \frac{s(\mathbf{e, t}) - \mu(\mathbf{s(\mathbf{t, I_{e}})})}{2\,\sigma(\mathbf{s(\mathbf{t, I_{e}})})}.
\end{equation}
\endgroup
\section{Trainable AS-norm}
\vspace{-0.1cm}
\subsection{Learning Methodology}
\vspace{-0.1cm}
Our TAS-norm is designed to be trained using in-domain datasets that also serve as impostors, enabling the AS-norm process to further fit to the in-domain. 
For the TAS-norm to be suitable for gradient-based learning, the scoring function $s(\cdot)$ should be differentiable, such as using Euclidean distance or cosine similarity.
Likewise, our TAS-norm is based on a variant of AS-norm1. The differentiability of this variant can be demonstrated as follows:
\begingroup\makeatletter\def\f@size{9.5}\check@mathfonts
\begin{equation}
\label{eqn:cohort_selections}
    \mathbf{s}(\mathbf{e, I_{e}}) = \text{top}_{K}\left(\mathbf{s}(\mathbf{e, I})\right), \,\,\, \mathbf{s}(\mathbf{t, I_{t}}) = \text{top}_{K}\left(\mathbf{s}(\mathbf{t, I})\right),
\end{equation}
\endgroup
where $\text{top}_{K}(\cdot)$ denotes top $K$ selection operation.
It is well known that this operation is differentiable \cite{diff_topk1, diff_topk2}.

During the learning process, there is a risk of including non-impostor speakers into the cohort, as the speakers of $\mathbf{e}$ and $\mathbf{t}$ present in $\mathbf{I}$. To address this issue, we draw inspiration from margin-based softmax approaches \cite{cosface, arcface}. We introduce a margin penalty $m$ to prevent non-impostor selection in constructing the cohort. 
Given the symmetrical nature of AS-norm in obtaining cohort scores for $\mathbf{e}$ and $\mathbf{t}$, we focus our explanation on the $\mathbf{e}$. Let the speaker of $\mathbf{e}$ be the $y_{c}$-th speaker label among total $C$ impostors. The $c$-th component of the penalized IEs set, denoted as $\bar{\mathbf{s}}(\mathbf{e, I_{e}})_{c}$, is adjusted as follows:
\vspace{-0.1cm}
\begingroup\makeatletter\def\f@size{9.5}\check@mathfonts
\begin{equation}\label{eqn:general_margin}
    \bar{\mathbf{s}}(\mathbf{e, I})_{c}
    =
    \begin{cases}
    \mathbf{s}(\mathbf{e, I})_{y_{c}} - m & \text{if $c = y_{c}$} \\
    \,\,\,\,\,\,\,\,\,\, \mathbf{s}(\mathbf{e, I})_{c} & \text{otherwise}
  \end{cases},
\end{equation}
\endgroup
where $m$ denotes the margin penalty. As a practical example, we consider cosine similarity, which is widely used as a scoring metric. We penalize the score by applying $m$ to the angle, similar to the angular additive margin (AAM) softmax \cite{arcface}, which can be represented as follows:
\vspace{-0.1cm}
\begingroup\makeatletter\def\f@size{9.5}\check@mathfonts
\begin{equation}\label{eqn:angle_margin}
    \bar{\mathbf{s}}(\mathbf{e, I})_{c}
    =
    \begin{cases}
    \text{cos}(\theta_{y_{c}} + m) & \text{if $c = y_{c}$} \\
    \,\,\,\,\,\,\,\,\,\, \text{cos}(\theta_{c}) & \text{otherwise}
  \end{cases},
\end{equation}
\endgroup
where $\theta_{c}$ is the angle between the $\mathbf{e}$ and $c$-th vector of $\mathbf{I}$.
The $\bar{\mathbf{s}}(\mathbf{t, I})$ is obtained with similar manner and we can get the normalized score of the TAS-norm as follows:
\begingroup\makeatletter\def\f@size{9.5}\check@mathfonts
\begin{eqnarray}
\label{eqn:TAS_norm}
    \begin{split}
    s_{tas}(\mathbf{e, t}) =
    \frac{s(\mathbf{e, t}) - \mu\left(\text{top}_{K}\left(
    \bar{\mathbf{s}}(\mathbf{e, I})\right)\right)}
    {2\,\sigma\left(\text{top}_{K}\left(\bar{\mathbf{s}}(\mathbf{e, I})\right)\right)} \\
    + \frac{s(\mathbf{e, t}) - \mu\left(\text{top}_{K}\left(
    \bar{\mathbf{s}}(\mathbf{t, I})\right)\right)}
    {2\,\sigma\left(\text{top}_{K}\left(\bar{\mathbf{s}}(\mathbf{t, I})\right)\right)}.
    \end{split}
\end{eqnarray}
\endgroup
The obtained normalized scores are optimized using CLLR function \cite{cllr, cllr_loss, wespeaker}, which is closely related to the detection cost function (DCF) \cite{dcf}, which is the widely used evaluation metric of the ASV task.
\vspace{-0.1cm}
\begingroup\makeatletter\def\f@size{9.5}\check@mathfonts
\begin{equation}
\label{eqn:CLLR}
    \mathcal{L}_{cllr} = \frac{1}{2\log2}
    \left( \frac{C_{tar}}{N_{tar}} + \frac{C_{non}}{N_{non}} \right),
\end{equation}
\endgroup
where $N_{tar}$ and $N_{non}$ are the number of target and non-target speakers in the mini-batch, respectively. 
The expected log costs of target score $C_{tar}$ and non-target score $C_{non}$ are computed as follows:
\vspace{-0.1cm}
\begingroup\makeatletter\def\f@size{9.5}\check@mathfonts
\begin{eqnarray}
\label{eqn:C_tar_and_C_non}
    C_{tar} = \sum_{N_{tar}}
    \left( \log \left(1 + \exp\left(s_{tas}^{tar}\right)\right) \right), \\
    C_{non} = \sum_{N_{non}}
    \left( \log \left(1 + \exp\left(s_{tas}^{non}\right)\right) \right),
\end{eqnarray}
\endgroup
where $s_{tas}^{tar}$ is a target score and $s_{tas}^{non}$ is a non-target score.

The STD of the normalized score distribution can be large enough to lower the $\mathcal{L}_{cllr}$, Therefore, the study that suggested $\mathcal{L}_{cllr}$ \cite{cllr_loss} used temperature scaling to adjust the score distribution to avoid this issue and ensure stable training. However, we utilize batch normalization \cite{bn} to normalized scores, which are more effective at normalization than temperature scaling.
To briefly introduce the differences from our previous study \cite{sp_cup24_hyu},
we physically removed non-impostors rather than using margin penalty. Additionally, we used negative log-likelihood loss function rather than using $\mathcal{L}_{cllr}$.

\subsection{Learnable Weight and Bias in Our Previous Study}
\vspace{-0.1cm}
In our previous study \cite{sp_cup24_hyu}, we leveraged LWBs and statistics obtained from AS-norm to learn the optimized mean and STD.
The normalized score of LWB-TAS-norm can be expressed by incorporating Eq. (\ref{eqn:TAS_norm}) and LWBs as follows:
\begingroup\makeatletter\def\f@size{9.5}\check@mathfonts
\begin{eqnarray}
\label{eqn:LWB-TAS_norm}
    \begin{split}
    s_{tas}(\mathbf{e, t}) =
    \frac{s(\mathbf{e, t}) - 
    \alpha \left( {w}_{\mu} \cdot
    \mu\left(\text{top}_{K}\left(
    \bar{\mathbf{s}}(\mathbf{e, I})\right)\right) +
    {b}_{\mu} \right)
    }
    {2
    \alpha \left( {w}_{\sigma} \cdot
    \sigma\left(\text{top}_{K}\left(\bar{\mathbf{s}}(\mathbf{e, I})\right)\right) +     {b}_{\sigma} \right)
    }  \\
    + \frac{s(\mathbf{e, t}) - 
    \alpha \left( {w}_{\mu} \cdot
    \mu\left(\text{top}_{K}\left(
    \bar{\mathbf{s}}(\mathbf{t, I})\right)\right)+
    {b}_{\mu} \right)
    }
    {2
    \alpha \left( {w}_{\sigma} \cdot
    \sigma\left(\text{top}_{K}\left(\bar{\mathbf{s}}(\mathbf{t, I})\right)\right) + 
    {b}_{\sigma} \right)
    },
    \end{split}
\end{eqnarray}
\endgroup
where scalars $w_{\mu}$, $w_{\sigma}$, $b_{\mu}$ and $b_{\sigma}$ are LWBs.
Also, $\alpha(\cdot)$ denotes the non-linear activation. In \cite{sp_cup24_hyu}, we used sigmoid function.

\subsection{Proposed Learnable Impostor Embeddings}
\vspace{-0.1cm}
The proposed LIE-TAS-norm is very simple in that we use IEs as a learnable parameter instead of using LWBs.
As used in the standard AS-norm, the initial values of the LIEs are set to be representative of the impostors, and then fine-tuned according to the learning methodology mentioned above.
\subsubsection{Sub-center Methodology}
Inspired by the AAM sub-center softmax \cite{sub-center-arcface}, we introduce $N_{sub}$ sub-centers for each impostor speaker. This allows the input embedding to select the most suitable center among these sub-centers. 
Consequently, the sub-center method facilitates the generation of more suitable IEs for each input utterance, thereby enhancing the adaptability and effectiveness of the LIE-TAS-norm.
Let $\mathcal{I} \in \mathcal{R}^{C \times N_{sub} \times V}$ represent the set of LIEs with sub-centers.
Applying the sub-center method to the $c$-th component of the penalized cohort in the example Eq. (\ref{eqn:angle_margin}), it is obtained as:
\vspace{-0.1cm}
\begingroup\makeatletter\def\f@size{9.5}\check@mathfonts
\begin{equation}\label{eqn:new_angle_margin}
   \bar{\mathbf{s}}(\mathbf{e, \mathcal{I}})_{c}
   =
   \begin{cases}
   f_{j} \left( \text{cos}(\theta_{y_{c},j} + m) \right) & \text{if $c = y_{c}$} \\
   \,\,\,\,\,\,\,\,\,\, f_{j} \left(  \text{cos}(\theta_{c,j}) \right)  & \text{otherwise}
 \end{cases},
\end{equation}
\endgroup
where $f_{j}(\cdot)$ denotes the selection function from the range of $j \in \{1,\cdots,J\}$. In our experiment, we use a \texttt{min} function.

\subsubsection{Auxiliary Impostor Classification Loss}
While we fine-tunes LIEs, a potential drawback emerges: LIEs might lose their original function as representatives of impostors during the fine-tuning process.
To ensure that LIEs do not forget their identity, we introduce $\mathcal{L}_{aic}$ aimed at regulating them by classifying impostors. The $\mathcal{L}_{aic}$ using Eq. (\ref{eqn:new_angle_margin}) is organized as follows:
\vspace{-0.1cm}
\begingroup\makeatletter\def\f@size{9.5}\check@mathfonts
\begin{equation}\label{eqn:loss_aic}
    \mathcal{L}_{aic} =-\frac{1}{M}\sum^{M}_{c=1}\log\frac{e^{\delta\cdot
    \bar{\mathbf{s}}(\mathbf{x, \mathcal{I}})_{y_{c}}
    }}{\sum_{c}^{C}{e^{\delta\cdot 
    \bar{\mathbf{s}}(\mathbf{x, \mathcal{I}})_{c}
    }}},
\end{equation}
\endgroup
where $\delta$ and $M$ are the scaling factor and mini-batch size, respectively. 
The total loss function is a summation of the scaled $\mathcal{L}_{aic}$ and $\mathcal{L}_{cllr}$.

\section{Experimental setup}
\vspace{-0.3cm}
\subsection{Dataset}
\vspace{-0.1cm}

We evaluated our TAS-norm methods using three linguistically diverse benchmarks. We utilized the VoxCeleb recipe \cite{vox1, vox2}, which comprised of multilingual datasets primarily composed of English speech, for the first benchmark. The second benchmark employed the CN-Celeb recipe \cite{cnceleb1, cnceleb2} for Chinese ASV system. Lastly, we implemented the SdSV 2020 challenge Task 2 (SdSV20 Task 2) protocol \cite{sdsv2020, deepmine} to assess Persian ASV system.
We used the progress subset and trials for evaluation, which were provided to participants along with the correct answers \cite{sdsv2020, sdsv20_hyu}.
Both Chinese and Persian benchmarks have multiple enrollment utterances per speaker. Therfore, we used the average of each enrollment embeddings as a representative for evaluation. 
\vspace{-0.1cm}
\subsection{Speaker Embedding Model}
We utilized the ECAPA-TDNN \cite{ecapa-tdnn} with basic channel size of 1,024 for speaker embedding model.
We set the speaker embedding size to 192-D, thus the total number of parameters was 14.7M.
AAM softmax loss \cite{arcface} with 0.2 margin penalty and scaling factor of 30 was utilized for training the ECAPA-TDNN.
The lengths of training waveform was randomly cropped into 2 s and it was transformed to 80-D log-mel spectrograms with a 25 ms frame length and 10 ms shift.
We pre-trained ECAPA-TDNN on VoxCeleb2 \textit{dev} \cite{vox2}, and then adapted ECAPA-TDNN for each dataset. 
During the pre-training, We applied speed perturbation augmentation \cite{speed_perturb1} by sampling a ratio from \{0.9, 1.1\}. These augmented waveforms were considered speakers who are different from the original ones \cite{speed_perturb1}. 
In addition, each waveform was reverberated with a OpenSLR simulated RIR dataset \cite{rir}, or augmented with noise in MUSAN dataset \cite{musan}.
The model was optimized using the Adam optimizer \cite{adam} with a mini-batch size was 200.
The initial learning rate was set to $10^{-3}$ and halved every 5 epochs, starting from epoch 60, during the entire 100 epoch training.
To obtain ECAPA-TDNNs fitted to each dataset, we initialized the optimizer and performed adaptation without augmentation.
Adaptation was performed in two steps. First, all layers except the last classification layer were frozen and only the last layer was trained for 20 epochs. After that, all layers were unfrozen and trained for 60 epochs.
The learning rate started at $10^{-4}$ and was decreased by 10\% every 5 epochs starting from 50 epoch of the total 80 epochs. 
\vspace{-0.1cm}

\input{tables/table2}
\vspace{-0.2cm}
\subsection{Details of the TAS-norm}
\vspace{-0.1cm}
We employed cosine similarity as $s(\cdot)$ for training and to evaluate the TAS-norm. 
Similar to \cite{fast_resnet, short_asv, mun22_interspeech}, mini-batch comprised 200 speakers, with two utterances randomly selected per speaker.
One serves as the enrollment utterance and the other as the test utterance. 
We extracted embeddings from the enrollment and test utterances, and then trained by generating 40,000 scores in each mini-batch, considering the all cases in every pair.
We ensured a input waveform lengths of 4 s by self-concatenating shorter utterances.
We used the Adam optimizer with an initial learning rate of $10^{-4}$, which was decreased by 10\% every epoch over a total of 20 epochs. 
Parameters were set as follows: cohort size $K = 400$, margin penalty $m = 0.5$, number of the sub-center $N_{sub}=2$, scaling factor $\delta$ = 30, and scaling factor of $\mathcal{L}_{aic}$ was $10^{-1}$.
\vspace{-0.1cm}
\section{Experimental Results and analysis}
\input{tables/table3}
As listed in Table I, we compared ASV performance of score normalization methods on VoxCeleb1-O testset \cite{vox1}. 
We used equal error rate (EER) and the minimum DCF (minDCF) with $10^{-2}$ target probability as evaluation metrics.
The results without any score normalization are marked as ”baseline”.
Compared with non-adaptive score normalization methods, S-norm performed best, and the two AS-norm variants slightly outperformed S-norm.
The LWB-TAS-norm did not show enhanced result over AS-norms. However, the proposed LIE-TAS-norm showed a significant performance improvement over AS-norms and LWB-TAS-norm. Specifically, the LIE-TAS-norm outperformed the AS-norm1 by relatively 4.11\% in EER and 10.62\% in minDCF.
The LIE-TAS-norm outperformed AS-norms on other benchmarks which comprised different languages as shown in Table II.
The results on CN-Celeb \cite{cnceleb1} showed similar trends to VoxCeleb trials \cite{vox1}.
The potential of TAS-norms was further demonstrated by the observation that LWB-TAS-norm performed reasonably on the SdSV20 Task 2 benchmark \cite{sdsv2020}.
Additionally, we adjusted the cohort size for both training and inference. As shown in Fig.\,1, the LIE-TAS-norm outperformed the AS-norm1, demonstrating its robustness across different cohort sizes.
\input{figures/fig1}
\input{tables/table4}
\input{figures/fig2}
\input{figures/fig3}

We conducted ablation studies of the LIE-TAS-norm with VoxCeleb benchmark.
We address two questions about the sub-center method. 
First, we examined the ASV performance change according to the number of sub-centers in Table III.
We observed the best performance with two sub-centers and achieved a 2.55\% and 3.43\% relative improvement in EER and minDCF, respectively, compared with not applying sub-center (one sub-center).
Consequently, this result confirmed that the sub-center methods allow the input embedding to align with a more suitable center, potentially improving the accuracy of impostor modeling and, in turn, the effectiveness of score normalization.
Second, we compared the ASV performance of using \texttt{max} and \texttt{min} functions, which were used for selecting centers in our sub-center method.
As shown in Fig.\,2(a), the performance gap between the \texttt{max} and \texttt{min} function widened as training progressed.
It demonstrates that selecting centers with lower similarity is beneficial.
Additionally, the ablation study on loss functions, shown in Fig.\,2(b), reveals that  $\mathcal{L}_{clrr}$ with the scaled $\mathcal{L}_{aic}$ performed better overall. When using $\mathcal{L}_{clrr}$ alone, we observed divergence starting in the middle of training.
However, $\mathcal{L}_{aic}$ alone led to a slight performance improvement compared with AS-norms as in Table I, because $\mathcal{L}_{aic}$ reduces the similarity between impostors, similar to the AAM softmax \cite{arcface}.
We conducted experiments on the length of training utterances.
As illustrated in Fig.\,3, we observed different performance trends across two benchmarks.
The VoxCeleb1 testset \cite{vox1} which primarily contains utterances longer than 7 s, performed better with longer training utterances. In contrast, the SdSV20 Task 2 testset \cite{sdsv2020}, with test utterances of 2 s or less, showed the best performance when training utterances were also 2 s long. This highlights TAS-norm's ability to optimize performance by adapting training utterance lengths to match the test environment.

\section{Conclusion}
We proposed TAS-norm, a novel ASV back-end system that utilizes learnable parameters.
The LIE-TAS-norm, which fine-tunes LIEs, demonstrated significant performance improvements over AS-norms across various ASV benchmarks.
We believe our TAS-norm will reinvigorate research in ASV back-end systems.
However, experiments in this study focused on in-domain datasets, future work will explore out-of-domain applications. 
Also, we will aim to enhance our LIE-TAS-norm by incorporating with other learnable parameter.
\vspace{-0.05cm}
\section{Acknowledgements}
\vspace{-0.05cm}
This work was supported by Institute of Information {\&} communications Technology Planning {\&} Evaluation (IITP) grant funded by the Korea government(MSIT) (No.\,2020-0-01373, Artificial Intelligence Graduate School Program (Hanyang University))
\bibliographystyle{IEEEtran}
\bibliography{refs}

\end{document}

%% file: tables/table2.tex
\begin{table}[t] \begin{center}
\captionsetup{singlelinecheck = false, justification=justified}
\caption{ASV performance using different score normalization methods on VoxCeleb1-O.
}
\vspace{-0.2cm}
\resizebox{0.71\columnwidth}{!}{
\begin{tabular} {cccc}
\toprule[1pt]
\textbf{Normalization} & \multirow{2}{*}{\textbf{Trainable}}& \multicolumn{2}{c}{\textbf{VoxCeleb1-O}} \\
\textbf{methods} && \textbf{EER} & \textbf{minDCF} \\ \midrule
baseline (no-norm) && 0.930 & 0.09223 \\
Z-norm &{\xmark}& 0.897 & 0.08814 \\
T-norm &{\xmark}& 0.897 & 0.08843 \\
S-norm &{\xmark}& 0.880 & 0.08612 \\
AT-norm &{\xmark}& 0.884 & 0.08753 \\
AS-norm1 &{\xmark}& 0.876 & 0.08539 \\
AS-norm2 &{\xmark}& 0.875 & 0.08588 \\
\midrule
LWB-TAS-norm &{\cmark}& 0.872 & 0.08535 \\
LIE-TAS-norm &{\cmark}& \textbf{0.840} & \textbf{0.07632} \\
\bottomrule[1pt]
\end{tabular}
}
\end{center}
\vspace{-0.8cm}
\end{table}

%% file: tables/table3.tex
\newcolumntype{C}{>{\centering\arraybackslash}p{4em}}
\begin{table*}[ht] \begin{center}
\caption{ASV performance comparison of adaptive score normalization methods on three benchmarks.}
\vspace{-0.3cm}
\resizebox{0.85\width}{!}{
\begin{tabularx}{0.85\textwidth} {cccccccccc}
\toprule[1pt]
\textbf{Normalization}& \multirow{2}{*}{\textbf{Trainable}} & \multicolumn{2}{c}{\textbf{VoxCeleb1-E}}& \multicolumn{2}{c}{\textbf{VoxCeleb1-H}}& \multicolumn{2}{c}{\textbf{CN-Celeb1 test}}& \multicolumn{2}{c}{\textbf{SdSV20 Task 2}}\\ 

\textbf{method} && EER(\%) & minDCF & EER(\%) & minDCF & EER(\%) & minDCF & EER(\%) & minDCF \\ \midrule
baseline && {1.158}& {0.13683} & {2.207}& {0.22495}& {7.529}& {0.42077} & {2.024}& {0.09247} \\
AS-norm1 & {\xmark} & {1.096}& {0.12722} & {2.133}& {0.21540}& {6.983}& {0.39883} & {1.814}& {0.08267} \\
AS-norm2 & {\xmark} & {1.085}& {0.12679} & {2.124}& {0.21493}& {6.872}& {0.40246} & {1.842}& {0.08547} \\
\midrule
LWB-TAS-norm & {\cmark} & {1.098}& {0.11890} & {2.147}& {0.21840}& {6.847} & {0.38296} & {1.765}& {0.07932} \\
LIE-TAS-norm & {\cmark} & \textbf{1.056}& \textbf{0.11593} & \textbf{1.956}& \textbf{0.19956}& \textbf{6.681} & \textbf{0.37484} & \textbf{1.742} & \textbf{0.07648} \\

\bottomrule[1pt]
\end{tabularx}
}
\end{center}
\vspace{-0.8cm}
\end{table*}

%% file: figures/fig1.tex
\begin{figure}[t]
  \centering
  \includegraphics[width=0.92\linewidth]{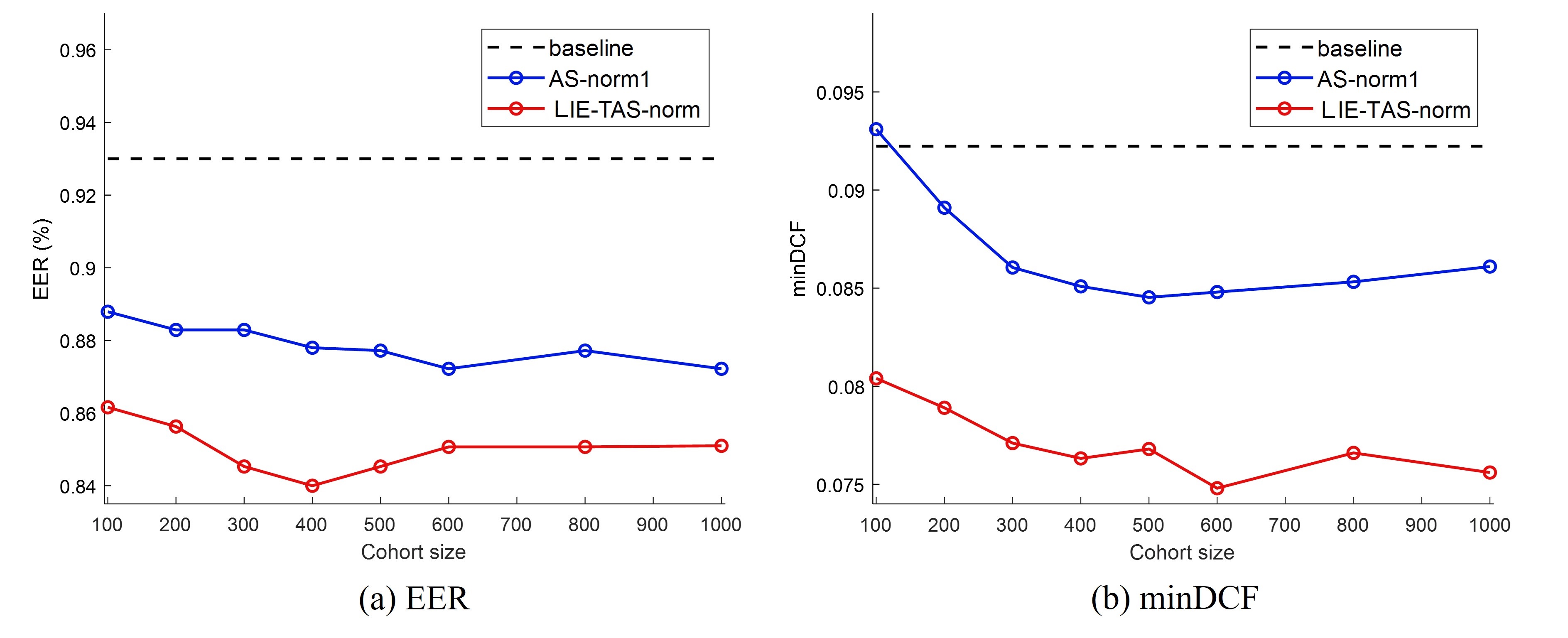}
  \vspace{-0.4cm}
  \caption{ASV performances on VoxCeleb1-O with different cohort size.}
  \label{figure1}
  \vspace{-0.5cm}
\end{figure}

%% file: tables/table4.tex
\begin{table}[t] \begin{center}
\captionsetup{singlelinecheck = false, justification=justified}
\caption{ASV performances on VoxCeleb1-O according to the number of sub-centers in the LIE-TAS-norm}
\vspace{-0.3cm}
\resizebox{0.78\columnwidth}{!}{
\begin{tabular} {cccccc}
\toprule[1pt]
\multirow{2}{*}{\textbf{Metric}} & \multicolumn{5}{c}{\textbf{Number of Sub-center}} \\ \cmidrule{2-6}
& 1 & 2 & 3 & 4 & 5 \\
\midrule
{EER(\%)} & 0.862 & \textbf{0.840} & 0.847 & 0.851 & 0.851 \\ \midrule
{minDCF} & 0.07903 & \textbf{0.07632} & 0.07719 & 0.07844 & 0.08026 \\ 
\bottomrule[1pt]
\end{tabular}
}
\end{center}
\vspace{-0.8cm}
\end{table}

%% file: figures/fig2.tex
\begin{figure}[t]
  \centering
  \includegraphics[width=0.92\linewidth]{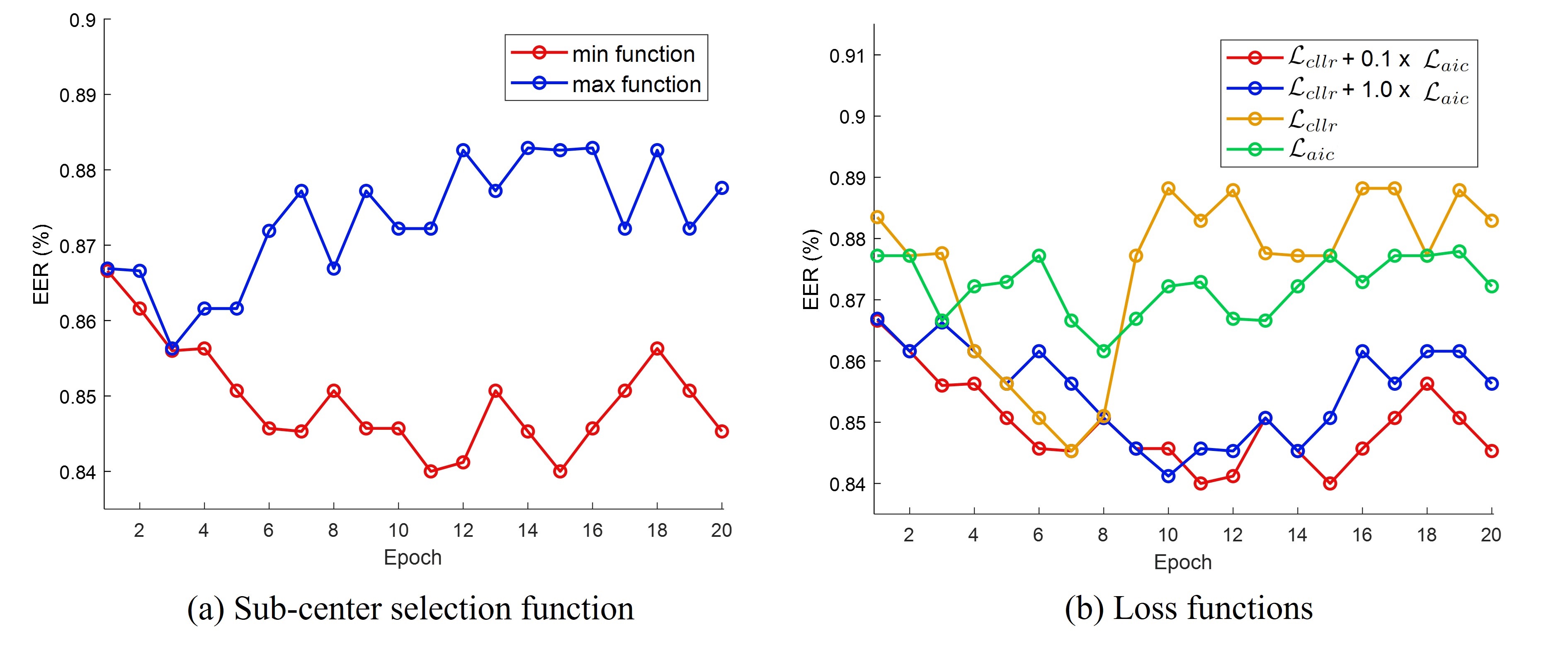}
  \vspace{-0.4cm}
  \caption{minDCF over training epochs: ablation studies on (a) selection functions of sub-center method and (b) loss functions.}
  \label{figure2}
  \vspace{-0.3cm}
\end{figure}

%% file: figures/fig3.tex
\begin{figure}[t]
  \centering
  \includegraphics[width=0.92\linewidth]{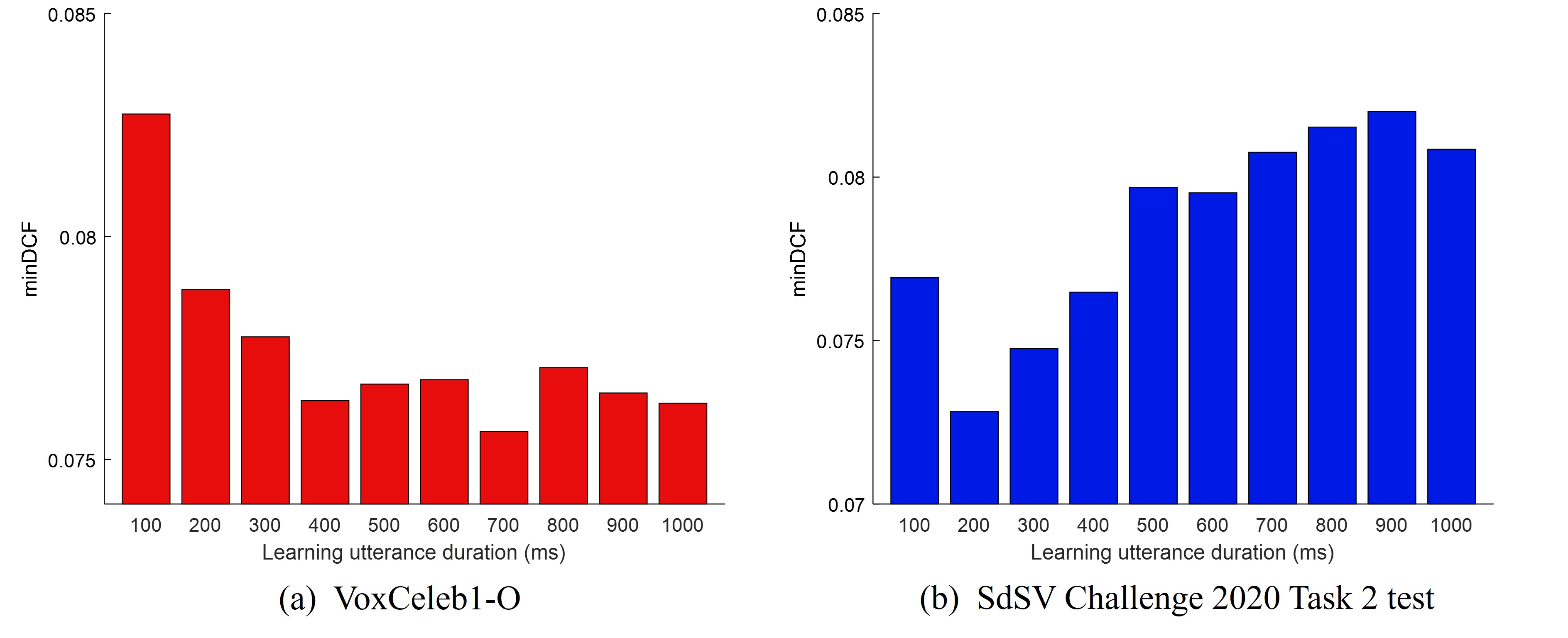}
  \vspace{-0.4cm}
  \caption{minDCF with different lengths of training utterances: (a) VoxCeleb1-O and (b) SdSV20 Task 2.}
  \label{figure3}
  \vspace{-0.6cm}
\end{figure}